\documentclass{article}
\usepackage{spconf,amsmath,graphicx,hyperref}
\usepackage{spconf,amsmath,amssymb,graphicx}
\usepackage{booktabs,multirow}
\usepackage{siunitx}
\usepackage{hyperref}
\hypersetup{hidelinks}
\usepackage{algorithm}
\usepackage{algorithmicx}
\usepackage{algpseudocode}
\usepackage{enumitem}
\usepackage{array}   
\usepackage{caption}
\usepackage{tabularx}

\title{PC-MCL: PATIENT-CONSISTENT MULTI-CYCLE LEARNING WITH MULTI-LABEL BIAS CORRECTION FOR RESPIRATORY SOUND CLASSIFICATION}
\name{Seung Gyu Jeong, Seong-Eun Kim$^{*}$ \thanks{$^{*}$Corresponding Author.} 
      \thanks{This work was supported by the National Research Foundation of Korea (NRF) funded by MSIT under Grants RS-2023-00208492 and RS-2024-00422599.}}
\address{Department of Applied AI, Seoul National University of Science and Technology, Seoul, South Korea\\
\ninept wa3229433@gmail.com, sekim@seoultech.ac.kr}

\begin{document}
\ninept
\maketitle

\begin{abstract}
Automated respiratory sound classification supports the diagnosis of pulmonary diseases. However, many deep models still rely on cycle-level analysis and suffer from patient-specific overfitting. We propose PC-MCL (Patient-Consistent Multi-Cycle Learning) to address these limitations by utilizing three key components: multi-cycle concatenation, a 3-label formulation, and a patient-matching auxiliary task. Our work resolves a multi-label distributional bias in respiratory sound classification, a critical issue inherent to applying multi-cycle concatenation with the conventional 2-label formulation (crackle, wheeze). This bias manifests as a systematic loss of normal signal information when normal and abnormal cycles are combined. Our proposed 3-label formulation (normal, crackle, wheeze) corrects this by preserving information from all constituent cycles in mixed samples. Furthermore, the patient-matching auxiliary task acts as a multi-task regularizer, encouraging the model to learn more robust features and improving generalization. On the ICBHI 2017 benchmark, PC-MCL achieves an ICBHI Score of 65.37\%, outperforming existing baselines. Ablation studies confirm that all three components are essential, working synergistically to improve the detection of abnormal respiratory events.
\end{abstract}


\section{Introduction}
Automated respiratory sound classification is emerging as a vital tool in digital healthcare, offering a low-cost, non-invasive method for screening and diagnosing lung diseases \cite{ref1,ref2}. By leveraging deep learning, such systems can analyze auscultation sounds, reducing the subjectivity and limited availability of human experts and thereby improving access to medical diagnostics \cite{ref3,ref4}. However, despite this promise, current deep learning approaches are constrained by two fundamental limitations that hinder their clinical impact. 

First, many methods adopt a single-cycle analysis paradigm, which overlooks temporal relations across cycles. Prior work has explored concatenation as self-augmentation by combining same-class cycles \cite{gairola2021respirenet,wang2022domain}, but the clinically relevant case of mixed pathological and normal segments remains underexplored. In real-world auscultation, abnormal sounds are often transient and intermittent \cite{andres2018respiratory, zulfiqar2021abnormal}. A single-cycle approach can easily miss these brief pathological events, failing to capture the broader temporal context in which they appear. The second limitation is patient-specific overfitting \cite{jeong2025patient}, where models learn superficial, patient-specific acoustics rather than the underlying pathology itself. This harms the model's ability to generalize to new, unseen patients.

\begin{figure*}[t!]
    \centering
    \includegraphics[width=\textwidth]{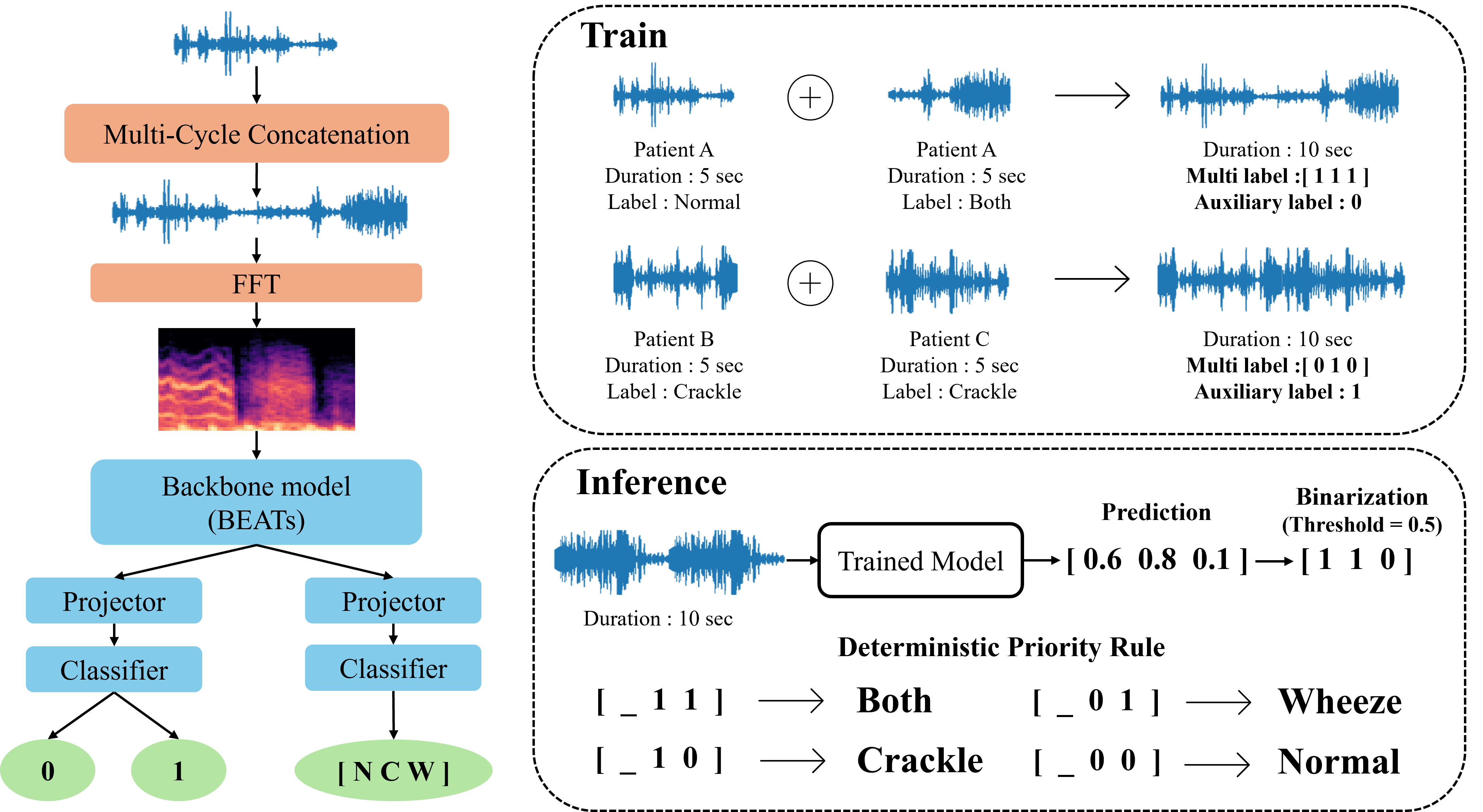} 
    \caption{Overview of the proposed PC-MCL framework. Individual respiratory cycles are combined via Multi-Cycle Concatenation, converted to a spectrogram, and fed into a shared backbone encoder. The resulting features are passed to two independent heads for the main pathology classification and the auxiliary patient-matching task.}
    \label{fig:framework}
\end{figure*}

To address the limitation of single-cycle analysis, we introduce the PC-MCL (Patient-Consistent Multi-Cycle Learning) framework, the overview of which is illustrated in Figure~\ref{fig:framework}. PC-MCL employs a multi-cycle concatenation strategy. By concatenating both normal and abnormal cycles, this approach mimics real-world auscultation, where pathological sounds are often transient and intermittent. It explicitly models the temporal context between different respiratory events, equipping the model with the ability to identify brief pathological patterns. 

However, we found that this can introduce a multi-label distributional bias in 2-label pathology settings (crackle and wheeze) \cite{noh2023rankmixup,yan2025simple}. 
To overcome this, we reformulate the task into a 3-label system (normal, crackle, wheeze), treating normal as an independent category. This new 3-label system ensures that when we combine cycles, the information from normal segments is preserved. 

Furthermore, to tackle patient-specific overfitting, PC-MCL incorporates a patient-matching auxiliary task where the model is simultaneously trained to identify if two cycles belong to the same patient. This multi-task learning setup acts as a regularizer. By learning patient-specific characteristics in addition to pathological sounds, the model is encouraged to develop a richer and more robust feature representation, which in turn improves generalization on the main classification task \cite{liebel2018auxiliary,shi2020auxiliary}.

\section{Proposed Method}
\label{sec:method}

\subsection{Multi-Cycle Concatenation and Bias Correction}
To overcome the limitation of single-cycle analysis, we propose a data augmentation strategy based on a multi-cycle concatenation. For each original training sample, we generate one augmented sample, including both \emph{self-augmentation} (same-class pairs) and \emph{cross-class mixing} (different-class pairs). To further increase diversity, these pairs are constructed from both intra- and cross-patient concatenation.

We form a composite input \(\tilde{x}\) of a fixed length \(T\) by concatenating two respiratory cycles, \(x^{(1)}\) and \(x^{(2)}\). Each cycle is first normalized to a length of \(T/2\) using repeat padding or center-cropping.
\[
\tilde{x} \;=\; \big[\, \tilde{x}^{(1)} \,\|\, \tilde{x}^{(2)} \,\big], \quad \text{where} \quad \tilde{x}^{(i)}=\mathrm{pad\_or\_crop}\big(x^{(i)}, T/2\big).
\]

A key challenge arises in labeling these combined cycles. Under the conventional 2-label [crackle, wheeze] system, a normal cycle is labeled \([0,0]\). When combined with a crackle cycle \([1,0]\), the resulting label for the mixed audio is also \([1,0]\). This approach erases the information that a normal sound was part of the mix, creating a significant training bias against the normal class in mixed-context scenarios. To resolve this information loss, we propose a new labeling scheme using a 3-dimensional vector [normal, crackle, wheeze]. When two cycles are combined, the new label \(y_{\text{main}}\) is generated by an element-wise logical OR (\(\lor\)) of their labels, \(y^{(1)}\) and \(y^{(2)}\).
\begin{equation}
y_{\text{main}}
\;=\;
y^{(1)} \,\lor\, y^{(2)}
\;=\;
\max\!\big(y^{(1)},y^{(2)}\big)
\end{equation}

For instance, combining a normal cycle \([1,0,0]\) and a crackle cycle \([0,1,0]\) now yields an informative label \([1,1,0]\). This method faithfully preserves the presence of all components, providing a more accurate training signal. The model is then trained using a standard multi-label binary cross-entropy loss on \(y_{\text{main}}\).

\begin{table*}[t!]
\centering
\captionsetup{skip=6pt}
\caption{Performance comparison with previous works that use the official split of the
ICBHI dataset (60-40\% split) and 4-class classification task. IN and AS refer to ImageNet \cite{deng2009imagenet} and AudioSet \cite{gemmeke2017audio}.}
\label{tab:performance_comparison}
\renewcommand{\arraystretch}{1.15}
\begin{tabularx}{\textwidth}{X l c c | c c c}
\toprule
Method & Backbone & Pretraining & Augmentation & Sp (\%) & Se (\%) & Score (\%) \\
\midrule

Ma et al. (LungRN+NL) \cite{ma2020lungrn+} & ResNet-NL & - & Mixup (2-label) & 63.20 & 41.32 & 52.26 \\
Gairola et al. (RespireNet) \cite{gairola2021respirenet} & ResNet34 & IN & Concat (same-class) & 72.30 & 40.10 & 56.20 \\
Wang et al. (Domain) \cite{wang2022domain} & ResNeSt & IN & Splicing (same-class) & 70.40 & 40.20 & 55.30 \\
Bae et al. (Patch-Mix CL) \cite{bae23b_interspeech} & AST & IN + AS & Patch-Mix & 81.66 & 43.07 & 62.37 \\
Kim et al. (RepAugment) \cite{kim2024repaugment} & AST & IN + AS & RepAugment & \textbf{82.47} & 40.55 & 61.51 \\
Jeong et al. (PAFA) \cite{jeong2025patient} & BEATs & AS & - & 82.05 & 47.63 & 64.84 \\

\cmidrule{1-7}
\textbf{AST + PC-MCL} & AST & IN + AS & Concat (same/cross-class) & 78.54\text{\scriptsize$\pm$1.87} & 
46.05\text{\scriptsize$\pm$1.62} & 
62.30\text{\scriptsize$\pm$0.50} \\
\textbf{BEATs + PC-MCL} & BEATs & AS & Concat (same/cross-class) & 79.04\text{\scriptsize$\pm$1.90}& $\textbf{51.71}$\text{\scriptsize$\pm$2.98}& $\textbf{65.37}$\text{\scriptsize$\pm$0.73}\\
\bottomrule
\end{tabularx}
\end{table*}

\begin{table}[t]
\centering
\caption{Performance Comparison with the Baseline.}
\label{tab:backbone}
\resizebox{0.48\textwidth}{!}{%
\begin{tabular}{lllccl}
\toprule
Model & Method  & Pretrain &  Sp (\%) & Se (\%) & Score (\%) \\
\midrule
\multirow{2}{*}{AST} 
 & CE   & \multirow{2}{*}{IN + AS}   & $77.14$\text{\scriptsize$\pm$5.43} & $41.97$\text{\scriptsize$\pm$5.04} & $59.55$\text{\scriptsize$\pm$0.50} \\
 & PC-MCL &                           & $\textbf{78.54}$\text{\scriptsize$\pm$1.87} & $\textbf{46.05}$\text{\scriptsize$\pm$1.62} & $\textbf{62.30}$\text{\scriptsize$\pm$0.50} \\
\midrule
\multirow{2}{*}{BEATs} 
 & CE   & \multirow{2}{*}{AS}        &
76.85\text{\scriptsize$\pm$1.88} & 
48.79\text{\scriptsize$\pm$1.72} & 
62.82\text{\scriptsize$\pm$0.62} \\
 & PC-MCL &                           & $\textbf{79.04}$\text{\scriptsize$\pm$1.90}& $\textbf{51.71}$\text{\scriptsize$\pm$2.98}& $\textbf{65.37}$\text{\scriptsize$\pm$0.73}\\
\bottomrule
\end{tabular}%
}
\end{table}

\subsection{Regularization via a Patient-Matching Auxiliary Task}
To mitigate overfitting to patient-specific acoustic characteristics, we employ a metadata-driven auxiliary task.

\begin{itemize}[leftmargin=*]
    \item \textbf{Architecture}: The architecture consists of a shared backbone encoder, $f_{\theta}$, which extracts a feature vector $z = f_{\theta}(x) \in \mathbb{R}^D$. This shared representation is passed to two independent, task-specific projection heads, $h_{\phi}^{\text{main}}$ and $h_{\phi}^{\text{aux}}$.
    
    \item \textbf{Main Task (Pathology Classification)}: The primary objective is to predict the 3-label pathology vector, $y_{\text{main}}$. The loss is the Binary Cross-Entropy with Logits Loss ($\mathcal{L}_{\text{BCE}}$):
    \begin{equation}
        \mathcal{L}_{\text{main}} = \mathcal{L}_{\text{BCE}}(h_{\phi}^{\text{main}}(z), y_{\text{main}})
    \end{equation}

    \item \textbf{Auxiliary Task (Patient-Matching)}: The auxiliary task predicts whether two concatenated cycles originate from the same patient ($y_{\text{aux}}=1$) or different patients ($y_{\text{aux}}=0$). We employ hard negative mining \cite{schroff2015facenet} by sampling negatives from different patients with the same pathology classes. This challenging task encourages the main classifier to focus on pathology-relevant cues while the auxiliary head captures patient-specific signatures. The auxiliary head uses Cross-Entropy loss:
    \begin{equation}
        \mathcal{L}_{\text{aux}} = \mathcal{L}_{\text{CE}}(h_{\phi}^{\text{aux}}(z), y_{\text{aux}})
    \end{equation}

    \item \textbf{Combined Loss}: The final training objective is a weighted sum of the main and auxiliary losses, where we set $\alpha=0.1$ based on a grid search on the validation set.
    \begin{equation}
        \mathcal{L}_{\text{total}} = \mathcal{L}_{\text{main}} + \alpha \cdot \mathcal{L}_{\text{aux}}
    \end{equation}
\end{itemize}

\subsection{Evaluation and Inference Strategy}

To evaluate our multi-label predictions against the single-label ICBHI benchmark, a deterministic conversion rule is required. We binarize the model's 3-dimensional probability output vector using a standard threshold of 0.5 to get a prediction vector. This vector is then converted into a single class based on a clinically driven priority rule: the co-occurrence of crackle and wheeze is mapped to the `both' class. Otherwise, the presence of crackle or wheeze maps to their respective classes. If no abnormal sounds are detected, the cycle is classified as `normal'. At inference time, each individual test cycle is padded or truncated to the target length and fed into the trained model. Although this introduces a slight domain shift between training (concatenated cycles) and inference (padded single cycles), we found this approach to be robust in practice.

\begin{table}[h!]
\centering
\caption{Component ablation study on the BEATs backbone. Concat refers to Multi-Cycle Concatenation, Multi to the 3 multi-label formulation, and PM to the Patient-Matching auxiliary task.}
\label{tab:ablate}
\begin{tabular}{ccccccc}
\toprule
Concat & Multi & PM & Sp (\%) & Se (\%) & Score (\%) \\
\midrule
 &  &  & 
76.85\text{\scriptsize$\pm$1.88} & 
48.79\text{\scriptsize$\pm$1.72} & 
62.82\text{\scriptsize$\pm$0.62} \\
& \checkmark &  & 
73.10\text{\scriptsize$\pm$2.47} & 
51.10\text{\scriptsize$\pm$2.97} & 
62.10\text{\scriptsize$\pm$0.64} \\
\checkmark &  &  & 
78.05\text{\scriptsize$\pm$2.49} & 
50.33\text{\scriptsize$\pm$2.93} & 
64.19\text{\scriptsize$\pm$0.31} \\
\checkmark & \checkmark &  & 
76.66\text{\scriptsize$\pm$2.93} & 
\textbf{51.91}\text{\scriptsize$\pm$2.25} & 
64.28\text{\scriptsize$\pm$0.62} \\
\checkmark & \checkmark & \checkmark & 
\textbf{79.04}\text{\scriptsize$\pm$1.90} & 
51.71\text{\scriptsize$\pm$2.98} & 
\textbf{65.37}\text{\scriptsize$\pm$0.73} \\
\bottomrule
\end{tabular}
\end{table}

\section{Experiments and Results}

\subsection{Experimental Setup}

All experiments are conducted on the ICBHI 2017 dataset using the official 60:40 split \cite{rocha2018alpha}. We adopt the official challenge metrics: Specificity ($S_p$), Sensitivity ($S_e$), and their mean, ICBHI Score (Score) \cite{rocha2018alpha}. To ensure statistical robustness, all reported results are the mean and standard deviation over five runs with different random seeds. All audio recordings are first resampled to 16,000 Hz. Following prior work \cite{beats, jeong2025patient}, the primary feature representation is a 128-dimensional Mel-Spectrogram, extracted using a 25ms window with a 10ms frame shift. For all experiments involving concatenation, the fixed input length for the network was set to 10 seconds.

\subsection{Main Results and Comparison}

We first compare our proposed framework against previous methods on the challenging 4-class classification task. As shown in Table~\ref{tab:performance_comparison}, our full framework achieves an ICBHI Score of 65.37\%, surpassing prior work and demonstrating the overall effectiveness of our proposed methodology. To further demonstrate the impact of our framework, we compare its performance against standard Cross-Entropy (CE) baselines on both AST and BEATs backbones. As presented in Table~\ref{tab:backbone}, our framework provides a consistent and significant performance uplift over the CE baseline for both architectures. Notably, the improvement is particularly significant in the clinically crucial metric of Sensitivity, with gains of 4.08 percentage points for AST and 2.92 percentage points for BEATs.

\begin{table}[t!]
\centering
\caption{Comparison of 2-label ([Crackle, Wheeze]) and our proposed 3-label ([Normal, Crackle, Wheeze]) formulation on BEATs.}
\label{tab:label_formulation}
\resizebox{0.48\textwidth}{!}{%
\begin{tabular}{lccc}
\toprule
Label Formulation & Sp (\%) & Se (\%) & Score (\%) \\
\midrule
2-label & $58.86 \pm 9.59$ & $\textbf{61.98} \pm 8.04$ & $60.42 \pm 1.31$ \\
\textbf{3-label (Ours)} & $\textbf{79.04} \pm 1.90$ & $51.71 \pm 2.98$ & $\textbf{65.37} \pm 0.73$ \\
\bottomrule
\end{tabular}%
}
\end{table}

\begin{table}[t!]
\centering
\caption{Patient-Matching auxiliary task scope on BEATs.}
\label{tab:hard}
\resizebox{0.37\textwidth}{!}{%
\begin{tabular}{lccc}
\toprule
Setting & Sp (\%) & Se (\%) & Score (\%) \\
\midrule
Base  & 
78.30\text{\scriptsize$\pm$4.20} & 
51.21\text{\scriptsize$\pm$3.77} & 
64.76\text{\scriptsize$\pm$0.26} \\
Hard  & 
\textbf{79.04}\text{\scriptsize$\pm$1.90} & 
\textbf{51.71}\text{\scriptsize$\pm$2.98} & 
\textbf{65.37}\text{\scriptsize$\pm$0.73} \\
\bottomrule
\end{tabular}%
}
\end{table}

\begin{figure*}[h!] 
    \centering
    \includegraphics[width=\textwidth]{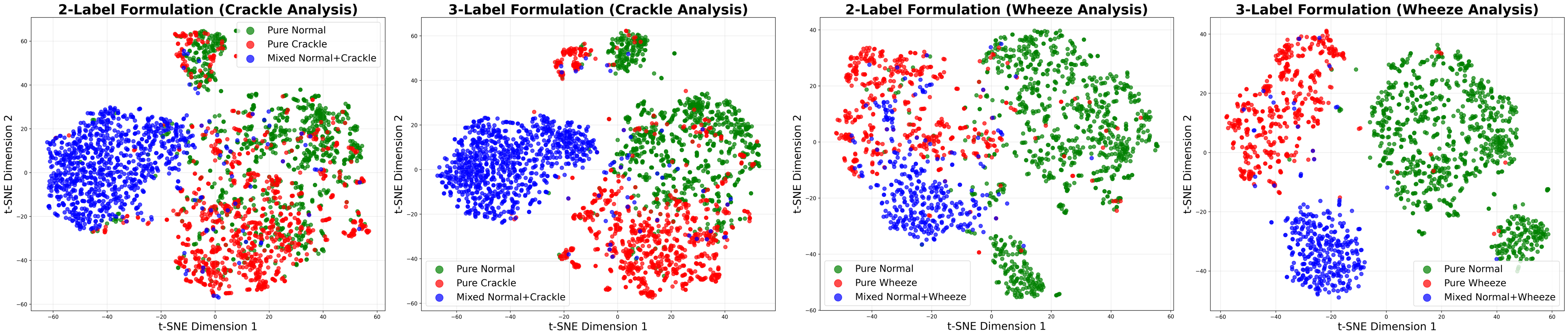} 
    \caption{Comparison of feature spaces learned by the conventional 2-label and our proposed 3-label models, visualized using t-SNE. The points represent three types of augmented samples: Pure Normal (green), Pure Abnormal (red),  and Mixed Normal+Abnormal (blue).}
    \label{fig:tsne_comparison}
\end{figure*}

\begin{figure*}[t!]
    \centering
    \includegraphics[width=\textwidth]{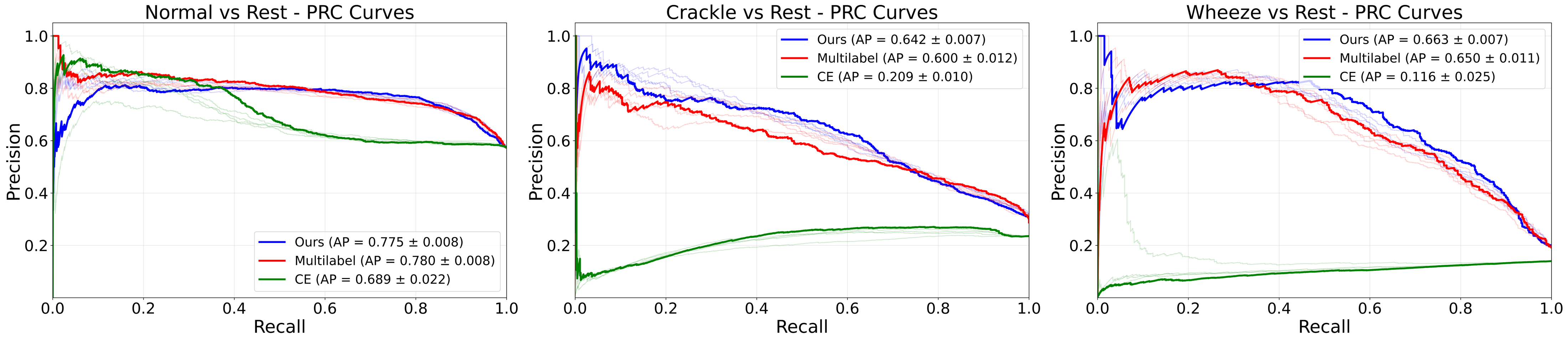} 
    
    \caption{Precision-Recall Curves for the three classes (Normal, Crackle, Wheeze) on the ICBHI test set. The curves compare the performance of the baseline CE model, an intermediate model with concatenation and multi-labeling (Multilabel), and our final proposed model (Ours). Average Precision (AP) scores, averaged over five runs, are shown in the legend for each model.}
    \label{fig:prc_curves}
\end{figure*}

\subsection{Ablation Studies}

\subsubsection{Component Analysis}

To validate the contribution of each component, we conducted a step-by-step ablation study, with results shown in Table~\ref{tab:ablate}. The analysis reveals the distinct and synergistic roles of our innovations. While introducing the Multi-label formulation alone primarily boosts Sensitivity (+2.31 points), Concatenation provides the largest increase to the overall Score (+1.37 points). The final addition of the Patient-Matching auxiliary task then acts as a powerful regularizer, improving Specificity and pushing the overall Score to its peak of 65.37\%. This confirms that each component is essential: Multi-labeling for sensitivity, Concatenation for general performance, and the PM task for final regularization.

\subsubsection{Analysis of Label Formulation}
A key design choice of our framework is the 3-label (normal, crackle, wheeze) formulation, where `normal' is treated as an independent label. This contrasts with the conventional 2-label (crackle, wheeze) setting, which implicitly defines `normal' as the complement of abnormal sounds. As shown in Table~\ref{tab:label_formulation}, the 2-label system, although achieving higher raw sensitivity, suffers from a substantial drop in specificity, resulting in a significantly lower ICBHI score.

This degradation directly stems from the multi-label distributional bias: concatenating a `normal' cycle with an `abnormal' one forces the entire sample to be labeled as purely abnormal in the 2-label setup, systematically erasing normal labels and biasing the model. To visually confirm this hypothesis, we analyzed the learned feature spaces using t-SNE \cite{maaten2008visualizing}, as shown in Figure~\ref{fig:tsne_comparison}. The 2-label model fails to distinguish between mixed (`normal+wheeze') and pure abnormal (`wheeze+wheeze') samples, mapping them to heavily overlapped regions. In contrast, our 3-label model successfully separates these sample types into distinct clusters. This provides direct visual evidence that our proposed 3-label formulation preserves crucial information from normal segments, learns a richer feature representation, and is therefore essential for mitigating the label distributional bias inherent in concatenation-based augmentation.

\subsubsection{Auxiliary Task - Negative Sampling Strategy}
We further analyzed the auxiliary patient-matching task by comparing a simple "Base" setting against a more challenging "Hard" setting. In the Hard setting, negative pairs are sampled from different patients with identical pathology profiles, forcing the model to disentangle pathology-related cues from patient-specific acoustic signatures. As shown in Table~\ref{tab:hard}, the Hard strategy improves the ICBHI score by 0.61 points, demonstrating that challenging negative samples act as a stronger regularizer, leading to more robust and generalizable representations.

\subsection{Threshold-Free Performance Analysis}

To evaluate our framework's pure discriminative power independent of any fixed decision threshold, we present a threshold-free analysis using Precision-Recall Curves (PRC), which are particularly informative for the imbalanced ICBHI dataset. Figure~\ref{fig:prc_curves} compares the per-class PRC of three models: the baseline Cross-Entropy (CE) model, an intermediate model using only our multi-label formulation, and our final proposed framework (Ours). The CE baseline fails to learn the abnormal classes, with extremely low Average Precision (AP) scores for Crackle (0.209) and Wheeze (0.116). Simply introducing the multi-label formulation yields a dramatic performance leap (AP scores of 0.600 and 0.650 for Crackle and Wheeze, respectively), demonstrating that modeling pathologies as co-occurring events is the most critical factor. Finally, our full framework (Ours) provides a further significant boost (AP to 0.642 and 0.663), confirming that the additional components synergistically enhance the model's underlying ability to discriminate pathological events across all decision thresholds.

\section{Conclusion}
In this paper, we introduced PC-MCL, a unified framework for robust respiratory sound classification. Our primary contribution is the identification and correction of multi-label distributional bias, a previously overlooked issue where multi-cycle concatenation skews label distributions. Our experiments confirmed that the proposed 3-label independent-normal formulation is critical for providing a balanced supervision signal and achieving a higher overall score than conventional 2-label approaches. Furthermore, we incorporate a patient-matching auxiliary task that serves as a powerful regularizer to mitigate patient-specific overfitting. Our comprehensive ablation studies confirm that all components work synergistically to achieve a high performance on the ICBHI benchmark.

\section{Acknowledgment}
The authors acknowledge the use of Google's Gemini Pro for assistance in the preparation and language refinement of this manuscript.




\bibliographystyle{IEEEbib}
\bibliography{strings,refs}

\end{document}